\documentclass[twocolumn,superscriptaddress,aps,preprintnumbers,amsmath,amssymb,prl,nofootinbib]{revtex4}

\usepackage{epsfig}
\usepackage{amsmath}
\usepackage{amssymb}
\usepackage{subfigure}
\usepackage{cancel}
\usepackage{textcomp}
\usepackage{calc}
\usepackage{graphicx}
\usepackage{dcolumn}
\usepackage{color}
\usepackage{xcolor}
\usepackage{pifont}
\usepackage{ulem}

\usepackage{MnSymbol}

\usepackage{hyperref}
\hypersetup{colorlinks=true,urlcolor=blue,linkcolor=red,citecolor=green!60!black}

\begin{document}


\title{From NANOGrav to LIGO with metastable cosmic strings}

\preprint{CERN-TH-2020-157}
\preprint{DESY 20-154}


\author{Wilfried Buchmuller}
\email{wilfried.buchmueller@desy.de}
\affiliation{Deutsches Elektronen Synchrotron DESY, 22607 Hamburg, Germany}

\author{Valerie Domcke}
\email{valerie.domcke@cern.ch}
\affiliation{Theoretical Physics Department, CERN, 1211 Geneva 23, Switzerland}
\affiliation{Institute of Physics, Laboratory for Particle Physics and Cosmology, EPFL, CH-1015, Lausanne, Switzerland}

\author{Kai Schmitz}
\email{kai.schmitz@cern.ch}
\affiliation{Theoretical Physics Department, CERN, 1211 Geneva 23, Switzerland}


\begin{abstract}
We interpret the recent NANOGrav results in terms of a stochastic gravitational wave background from metastable cosmic strings. The observed amplitude of a stochastic signal can be translated into a range for the cosmic string tension and the mass of magnetic monopoles arising in theories of grand unification. In a sizable part of the parameter space, this interpretation predicts a large stochastic gravitational wave signal in the frequency band of ground-based interferometers, which can be probed in the very near future. We confront these results with predictions from successful inflation, leptogenesis and dark matter from the spontaneous breaking of a gauged $B$$-$$L$ symmetry.
\end{abstract}


\date{\today}
\maketitle


\noindent\textbf{Introduction}

The direct observation of gravitational waves (GWs) generated by
merging black holes \cite{Abbott:2016blz,Abbott:2016nmj,Abbott:2017vtc}
has led to an increasing interest in further explorations of the
GW spectrum. Astrophysical sources can lead to a stochastic
gravitational background (SGWB) over a wide range of frequencies, and
the ultimate hope is the detection of a SGWB of cosmological origin.
So far, transient merger events have been observed at frequencies around
$100~\text{Hz}$. Moreover, stringent upper bounds on a SGWB have been
obtained by pulsar timing array (PTA) experiments which are sensitive
to frequencies around $f_\text{yr} = 1/\text{yr}$. Over the past years the
European Timing Array (EPTA) \cite{Shannon:2015ect}, the Parkes
Pulsar Timing Array (PPTA) \cite{Kerr:2020qdo} and the North
American Nanohertz Observatory for Gravitational Waves (NANOGrav)
\cite{Arzoumanian:2018saf} have reached upper bounds on the amplitude
$h^2\Omega_\text{gw}(1/\text{yr})$ of order $10^{-9}$.

Searching for an isotropic SGWB, the NANOGrav Collaboration has
recently reported strong evidence of a stochastic process in their lowest frequency bins, which can
be modeled as a power-law with common amplitude and slope across all
pulsars \cite{Arzoumanian:2020vkk}. The amplitude of the signal is of
the order of the previously obtained upper bounds. The current data is not
conclusive with respect to a quadrupolar spatial correlation
and therefore the discovery of a SGWB cannot be claimed.
Nevertheless, the result of the analysis is very intriguing, and the
NANOGrav Collaboration finds that the signal is consistent, within
$2\sigma$ of a Bayesian analysis, with a SGWB from supermassive black
hole binaries,
the expected dominant astrophysical source
at frequencies around $1/\text{yr}$ \cite{ Rajagopal:1994zj,Phinney:2001di}.

There are also cosmological interpretations of the NANOGrav
results. Examples are the formation of primordial black holes from
high-amplitude curvature perturbations during inflation
\cite{Vaskonen:2020lbd,DeLuca:2020agl} or dark sector phase transitions~\cite{Nakai:2020oit}. Another prominent possibility is cosmic
strings formed in a $\text{U}(1)$ symmetry-breaking phase transition
in the early universe \cite{Kibble:1976sj,Jeannerot:2003qv}. Indeed,
it has been demonstrated that GWs from a network of stable strings
with an amplitude $h^2\Omega_\text{gw}(1/\text{yr}) \sim 10^{-9}$ can account for the
NANOGrav stochastic background \cite{Ellis:2020ena,Blasi:2020mfx}.
This signal is too small to be observed by Virgo~\cite{AdvVirgo}, LIGO~\cite{aLIGO_era_first} and KAGRA~\cite{Akutsu:2018axf}
but will be probed by LISA \cite{Auclair:2019wcv} and other planned GW observatories.

In this Letter we study a further possibility, metastable cosmic
strings. Recently, it has been shown that GWs emitted from a metastable cosmic string network can
probe the seesaw mechanism of neutrino physics and high-scale
leptogenesis \cite{Dror:2019syi}
as well as the energy scale of grand unification
\cite{Buchmuller:2019gfy,King:2020hyd}.
Such metastable cosmic strings arise when connecting hybrid inflation, high-scale leptogenesis and dark
matter with gravitational waves through $\text{U}(1)_{B-L}$ breaking 
in a cosmological phase transition \cite{Buchmuller:2012wn,Buchmuller:2013lra}.  
Here $B\!-\!L$ denotes the difference of baryon number and lepton
number, and the product of $\text{U}(1)_{B-L}$ and the Standard Model gauge group
is embedded into the GUT group SO(10). 
If the $\text{U}(1)_{B-L}$ cosmic strings are not protected by an additional unbroken discrete symmetry,
this embedding leads to the existence of magnetic monopoles, allowing the cosmic strings to 
decay via the Schwinger
production of monopole-antimonopole pairs with a rate per string unit length of~\cite{Leblond:2009fq, Monin:2008mp,Monin:2009ch}\footnote{{Due to their large mass, these monopoles can only be created prior to the final 60 e-folds of cosmic inflation or on the cosmic strings. In both cases, there are no remnant magnetic monopoles in our present Universe.}}
\begin{equation}
  \Gamma_d = \frac{\mu}{2 \pi} \exp\left( - \pi \kappa \right) \ ,\quad
  \kappa = \frac{m^2}{\mu} \ ,
\end{equation}
where $m \sim v_\text{GUT}$ is the
monopole mass and  $\mu \sim v_{B-L}^2$ is the string tension.
Here $v_\text{GUT}$ and $v_{B-L}$ are the scales of SO(10) and
$\text{U}(1)_{B-L}$ symmetry breaking, respectively.

At frequencies around $100~\text{Hz}$ the model of \cite{Buchmuller:2012wn} predicts a GW
amplitude close to the present upper bound found by the LIGO/Virgo collaboration, and upper
bounds on a SGWB by PTA experiments lead to an upper bound on the ratio
$\kappa$ and therefore on the monopole mass
\cite{Buchmuller:2019gfy}.
With the new NANOGrav data \cite{Arzoumanian:2020vkk}, 
$\kappa$ and hence the scale of grand unification $v_{\rm GUT}$ can now be
determined.


\medskip\noindent\textbf{GWs from metastable cosmic strings}

We briefly review the calculation of the stochastic gravitational wave background arising from metastable cosmic strings~\cite{Buchmuller:2019gfy}. 
The present-day GW spectrum can be expressed as~\cite{Auclair:2019wcv}
\begin{align}
\Omega_\text{gw}(f) = \frac{\partial \rho_\text{gw}(f)}{\rho_c \partial \ln f}= \frac{8 \pi f (G \mu)^2}{3 H_0^2} \sum_{n = 1}^\infty C_n(f) \, P_n \,,
\label{eq:Omega}
\end{align}
where $\rho_\text{gw}$ denotes the GW energy density, $\rho_c$ is the critical energy density of the universe, $G \mu$ denotes the dimensionless string tension with the gravitational constant $G= 6.7 \cdot 10^{-39}~\text{GeV}^{-2}$, $H_0 = 100 \,h\,\textrm{km}/\textrm{s}/\textrm{Mpc}$ is today's Hubble parameter, $P_n \simeq 50/\zeta[4/3] \,n^{-4/3}$ is the power spectrum of GWs emitted by the $n^{\rm th}$ harmonic of a cosmic string loop\footnote{
Here we focus on cusps as the main source of GW emission, kinks and kink-kink collisions yield a different $\cal{O}$(1) factor in both
the argument of the $\zeta$ function and the power of $n$ in $P_n$.
}, and $C_n(f)$ indicates the number of loops emitting GWs that are observed at a given frequency $f$,
\begin{align}
\label{eq:Cn}
C_n(f) = \frac{2 n}{f^2} \int_{z_\text{min}}^{z_\text{max}}dz\:\frac{\mathcal{N}\left(\ell\left(z\right),\,t\left(z\right)\right)}{H\left(z\right)(1 + z)^6} \,,
\end{align}
which is a function of the number density of cosmic string loops $\mathcal{N}(\ell,t)$, with $\ell = 2n/((1 + z) f)$, selecting the loops that contribute to the spectrum at frequency $f$ today.
Modeling the evolution and GW emission of a cosmic string network is a challenging task, resulting in several competing models for the loop number density in the literature (see~\cite{Auclair:2019wcv} for an overview). \{For concreteness, we will base our analysis on the Blanco-Pillado--Olum--Shlaer (BOS) model~\cite{Blanco-Pillado:2013qja} and fix the cosmic string loop size to $\alpha= \ell/H = 0.1$ at formation.
This roughly corresponds to the peak in the distribution of $\alpha$ values found in~\cite{Blanco-Pillado:2013qja}.
The peak itself has a width of less than an order of magnitude, which translates into an uncertainty in the GW signal of less than half an order of magnitude~\cite{Auclair:2019wcv}.
We also note that the assumption of fixed $\alpha$ is relaxed in~\cite{Blasi:2020mfx}, which scans over a larger range of $\alpha$ values.%
\footnote{{Because the expression in Eq.~\eqref{eq:nr} is only valid for $\alpha = 0.1$, the analysis in~\cite{Blasi:2020mfx} employs the analytical velocity-dependent one-scale (VOS) model~\cite{Martins:1995tg,Martins:1996jp,Martins:2000cs} instead of the BOS model.}}
For loops generated and decaying during the radiation-dominated era, this yields in particular~\cite{Blanco-Pillado:2013qja,Auclair:2019wcv}
\begin{align}
\mathcal{N}_r(\ell,t) & = \frac{0.18}{t^{3/2} (\ell + \Gamma G \mu t)^{5/2}}  \,, \label{eq:nr}
\end{align}
where $\Gamma \simeq 50$ parametrizes the cosmic string decay rate into GWs, $\dot \ell = - \Gamma G \mu$. This yields the dominant contribution to the GW spectrum in most of the parameter range of interest, but in our numerical computation of the spectrum we also include the loops created and/or decaying in the matter dominated era. 
The integration range in Eq.~\eqref{eq:Cn} accounts for the lifetime of the cosmic string network, from the formation at $z_\text{max}$ 
until their decay at $z_\text{min}$ when the decay rate of a string loop with average length  equals the Hubble rate~\cite{Leblond:2009fq},\footnote{In the U$(1)_{B-L}$ model~\cite{Buchmuller:2012wn,Buchmuller:2019gfy}, the formation time of the cosmic string network coincides with the reheating epoch after inflation, {i.e.\ $z_\text{max} \simeq T_\text{rh}/(2.7~\text{K})$, with $T_\text{rh}$ denoting the reheating temperature. In the viable parameter space of \cite{Buchmuller:2019gfy}, the latter takes values of $10^8 < T_\text{rh} < 10^{10}$~GeV, determined by the decay of $B$$-$$L$ Higgs fields and right-handed neutrinos. For such high reheating temperatures, the details of the reheating process and the string formation only impact the GW spectrum at very high frequencies beyond the range discussed here~\cite{Auclair:2019wcv}. }}
\begin{equation}
z_\text{min} = \left( \frac{70}{H_0}\right)^{1/2} \left( \Gamma  \; \Gamma_d  \; G \mu \right)^{1/4} \,.
\label{eq:zmin}
\end{equation}
For cosmic string loops formed and emitting GWs in the radiation dominated era, this results in an approximately scale invariant GW spectrum.
The finite lifetime of the cosmic strings leads to a fall-off $\propto f^{3/2}$ of this spectrum at small frequencies $f < f_*$ with~\cite{Buchmuller:2019gfy}
\begin{align}
\label{eq:fs_ana}
f_* \simeq 4.4 \times 10^{-8} \, \text{Hz} \:\frac{e^{- \pi \kappa/4}}{e^{- 16 \pi}} \left(\frac{ 10^{-7}}{G \mu }\right)^{1/2} \,,
\end{align}
see Fig.~\ref{fig:spectrum} for some examples of GW spectra for different values of the two dimensionless model parameters $G\mu$ and $\kappa$.

For the numerical evaluation of Eq.~\eqref{eq:Omega}, we refine the analysis of Ref.~\cite{Buchmuller:2019gfy} by resumming the first 20,000 modes and taking into account the changes in the number of effective degrees of freedom in the thermal bath (see also \cite{Gouttenoire:2019kij}). Our final results prove rather insensitive to both these refinements. Approximating ${\cal N} \simeq {\cal N}_r$, we can extract the $n$-dependence of $C_n P_n$ analytically if $\ell$ is much smaller or larger than $\Gamma G \mu t$. As discussed in Ref.~\cite{Buchmuller:2019gfy}, this distinction corresponds to the $f^{3/2}$ slope and the plateau regime. For the former, we find $C_n P_n \propto n^{-17/3}$, such that the resummation yields $\Omega_\text{gw} = \zeta(17/3) \, \Omega_\text{gw}^{(1)} \simeq 1.02 \,  \Omega_\text{gw}^{(1)}$, with $\Omega_\text{gw}^{(1)}$ denoting the result for $n=1$. For the plateau value, we instead obtain a factor $\zeta(4/3) \simeq 3.6$, which implies an ${\cal O}(1)$ correction.

For the evolution of the degrees of freedom we use the results of~\cite{Saikawa:2020swg} for the SM degrees of freedom and moreover include supersymmetric degrees of freedom at a threshold value of 2~TeV. This does not impact the predictions in the NANOGrav frequency range.


\medskip\noindent\textbf{Explaining the NANOGrav results}

\begin{figure}[t]
\begin{center}
\includegraphics[width=0.48\textwidth]{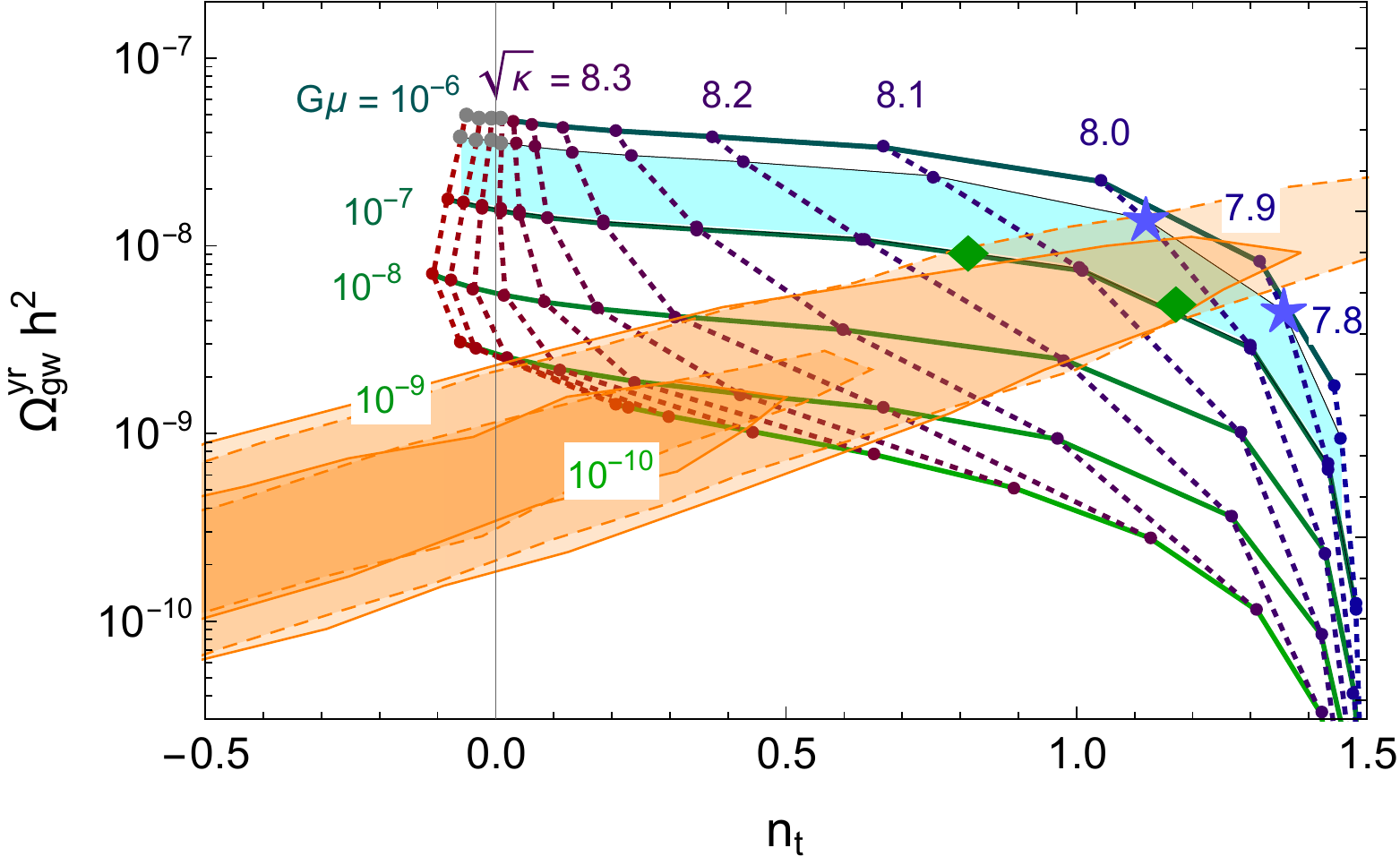}
\caption{Gravitational wave signals from metastable cosmic strings compared to the NANOGrav observations for different values of the string tension $G\mu$ and the hierarchy between the GUT and U(1) breaking scale $\kappa$. 
The solid coloured lines indicate fixed values of $G\mu = 10^{-10},..,10^{-6}$, the dotted lines indicate contours of $\sqrt{\kappa} = 7.8, 7.9,..,9$.
The orange region with the solid (dashed) contours show the 68$\%$ and 95$\%$ regions reported by NANOGrav when performing a fit to the first 5 frequency bins (performing a fit with a broken power law). 
}
\label{fig:summary}
\end{center}
\end{figure}

\begin{figure}[t]
\begin{center}
\vspace{-3.7mm}
  \includegraphics[width=0.48\textwidth]{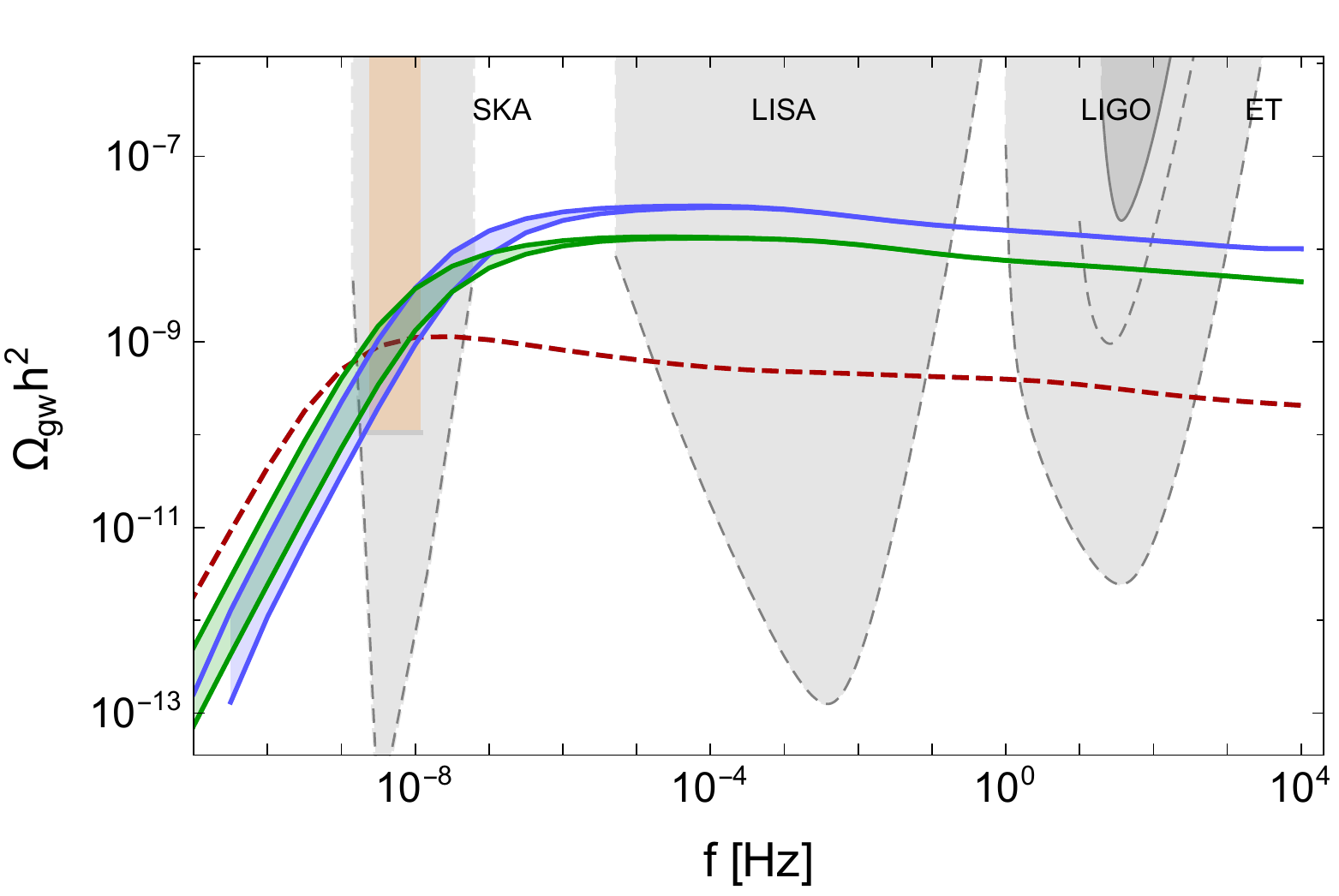}
\caption{Gravitational wave spectra from metastable cosmic strings explaining the NANOGrav excess (at 2$\sigma$ CL). The colored blue (green) region accounts for successful inflation, baryogenesis and dark matter~\cite{Buchmuller:2012wn,Buchmuller:2019gfy} for the maximal (minimal) allowed value of $G\mu$ with $\sqrt{\kappa} = 7.9..8.0$ (8.05..8.15), corresponding to the $\star$ ($\filleddiamond$) markers in Fig.~\ref{fig:summary}. {For reference, the dashed red line shows the spectrum for stable cosmic strings for the best-fit value $G\mu = 10^{-10}$~\cite{Ellis:2020ena,Blasi:2020mfx}, this curve is degenerate with the corresponding curve with $\sqrt{\kappa} = 9$.}
The (lighter) gray-shaded areas indicate the sensitivities of the (planned) experiments SKA~\cite{Smits:2008cf}, LISA~\cite{Audley:2017drz}, LIGO/Virgo~\cite{LIGOScientific:2019vic} and ET~\cite{ET}. The orange band indicates the frequency range of NANOGrav.}
\label{fig:spectrum}
\end{center}
\end{figure}

We now proceed to comparing the GW signal predicted by metastable cosmic strings to the recent NANOGrav results~\cite{Arzoumanian:2020vkk}, which  constrain the amplitude and slope of a stochastic process. Expressing the dimensionless characteristic strain as
$h_c = A (f/f_\text{yr})^\alpha$ with the reference frequency $f_\text{yr}=32$~nHz, the amplitude of the SGWB is obtained as
\begin{align}
 \Omega_\text{gw}(f)  = \frac{2 \pi^2 f_\text{yr}^2 A^2 }{3 H_0^2}  \left( \frac{f}{f_\text{yr}} \right)^{2 \alpha + 2} \equiv \Omega_\text{gw}^\text{yr} \left( \frac{f}{f_\text{yr}} \right)^{n_t}\,.
\end{align}
{This allows us to directly translate the one and two sigma confidence intervals given in~\cite{Arzoumanian:2020vkk} into the $\Omega_\text{gw}^\text{yr} - n_t$ plane, as depicted by the orange shaded region in Fig.~\ref{fig:summary}. To compute the theory prediction for a given parameter point, we evaluate $\Omega_\text{gw}(f)$ at the five frequencies corresponding to the five frequency bins used in the analysis of ~\cite{Arzoumanian:2020vkk}, and then extract the parameters $\Omega_\text{gw}^\text{yr}$ and $n_t$ by performing a least squares power-law fit. This procedure ensures that both theory and experimental data are evaluated in the same frequency range,  $f = 2.4..12$~nHz, which lies somewhat below the reference frequency $f_\text{yr}=32$~nHz.  This rather simple procedure is sufficient for our purpose, since the predicted spectral shape can be reasonably well approximated by a power law in this frequency range, as can be seen from Fig.~\ref{fig:spectrum}.}



In Fig.~\ref{fig:summary}, we compare these predictions from metastable cosmic strings (mesh of solid and dotted curves) with the constraints on the amplitude and tilt from \cite{Arzoumanian:2020vkk} (orange shaded region). We vary $G\mu$ from the lowest value capable of explaining the NANOGrav results at 2~sigma, $G\mu \simeq 10^{-10}$, to the largest value compatible with the constraints from LIGO/Virgo~\cite{LIGOScientific:2019vic}, $G\mu \simeq 10^{-6}$. The CMB constraint $G\mu < 1.3 \times 10^{-7}$~\cite{Ade:2015xua} only applies to cosmic strings with a life-time exceeding CMB decoupling, corresponding to $\sqrt{\kappa} \gtrsim 8.6$ (indicated by gray points in the upper left corner). For each value of $G\mu$, we consider the range $\sqrt{\kappa} = 7.8..9.0$; 
smaller values lead to an unobservably small spectrum at nHz frequencies, while all values $\sqrt{\kappa} \gtrsim 9$ quickly converge towards the result for stable cosmic strings, see~\cite{Ellis:2020ena,Blasi:2020mfx}. Contours of constant $G\mu$ ($\kappa$) are indicated by solid (dotted) lines in Fig.~\ref{fig:summary}.

The cyan shaded band in Fig.~\ref{fig:summary} indicates the prediction from $B$$-$$L$ breaking in the early Universe~\cite{Buchmuller:2012wn,Buchmuller:2019gfy}. Remarkably the predicted GW signal at nHz frequencies is compatible with the NANOGrav results at 2 sigma. 

It is intriguing that the values of the cosmic string tension $G\mu$ found in the context of metastable cosmic strings can be significantly larger than the values found for stable cosmic strings~\cite{Ellis:2020ena,Blasi:2020mfx}, implying the possibility of observing this signal with the existing ground-based detectors Virgo, LIGO and KAGRA. The reason for this is twofold. Firstly, the finite lifetime of the cosmic strings leads to a suppression of the low-frequency spectrum, implying a blue tilt of the GW spectrum between the {range of} PTAs and the frequency band of ground-based interferometers. In particular, the production of GWs after matter-radiation equality is suppressed, which for stable cosmic strings leads to a {mild} enhancement at low frequencies, see e.g.\ the {dashed} red curve in Fig.~\ref{fig:spectrum}. Secondly, the NANOGrav data exhibit a sizable correlation between the amplitude and tilt of the spectrum, allowing for larger amplitudes for positive values of $n_t$.


\medskip\noindent\textbf{Discussion}

In the model of cosmological $\text{U}(1)_{B-L}$ breaking \cite{Buchmuller:2012wn,Buchmuller:2019gfy},
successful inflation, leptogenesis and dark matter restrict the
allowed values of $G\mu$ to a narrow band around $G\mu \sim 3\times
10^{-7}$, depicted by the cyan region in Fig.~\ref{fig:summary}. Interpreting the 
NANOGrav results as originating from a metastable cosmic string network
determines the ratio between the GUT and the $B$$-$$L$ breaking scales to lie around $\sqrt{\kappa} \simeq 8$,
excluding stable cosmic strings. 
More precisely, the predictions of \cite{Buchmuller:2012wn,Buchmuller:2019gfy} are 
consistent with the recent NANOGrav results in the range from 
$G\mu = 1.0 \times 10^{-7}$,
with $\sqrt{\kappa} = 8.05..8.15$, to  $G\mu = 5.6 \times 10^{-7}$, with
$\sqrt{\kappa} = {7.9 .. 8.0}$. The corresponding values of $B\!-\!L$
breaking scales and monopoles masses are $v_{B-L} = 3.0 \times
10^{15}~\text{GeV}$, with $m {\simeq 3.2}\times 10^{16}~\text{GeV}$ and
$v_{B-L} = 5.8 \times 10^{15}~\text{GeV}$,
with $m = {(7.2 .. 7.3)}\times 10^{16}~\text{GeV}$, respectively.
The precise connection between GUT symmetry breaking, inflation and
$\text{U}(1)_{B-L}$ is a challenging theoretical question.\footnote{Determining the ${\cal O}(1)$ factors between
the ratio of monopole mass and string tension (parametrized by $\sqrt{\kappa}$)  and the ratio of 
the underlying scales $v_\text{GUT}/v_{B-L}$ requires a careful and consistent treatment of both
types of topological defects under consideration of the gauge coupling and the symmetry breaking potentials. }

A second important outcome of our analysis are the expectations for ground-based GW interferometers.
In Fig.~\ref{fig:spectrum} the GW spectrum is shown for the upper and the
lower boundary of the range in $G\mu$ that is predicted by the considered
$\text{U}(1)_{B-L}$ model \cite{Buchmuller:2012wn}. 
The prediction of this
model will be probed by Advanced LIGO
\cite{LIGOScientific:2019vic}.\footnote{On the contrary, interpreting the NANOGrav signal as originating from stable cosmic strings forces $G\mu$ to values too low to be observed by current ground-based GW interferometers~\cite{Ellis:2020ena}.}  The observation of a SGWB with PTA
experiments as well as with LIGO would give stunning support for
grand unified theories, with important implications for
inflation, baryogenesis and dark matter \cite{Buchmuller:2019gfy}.

An improved determination of the tilt of the spectrum at PTA frequencies together with upcoming results on SGWBs at LIGO frequencies will soon rule out or further support the model presented here.  This encourages further refinements of the analysis, e.g.\ going beyond the instantaneous decay approximation for the cosmic string network and taking into account the dynamics of cosmic string decay induced by monopole formation, which may lead to an additional GW contribution~\cite{Vilenkin:1982hm,Leblond:2009fq}. One may also consider relaxing some of the  model-building assumptions within the model of cosmological $\text{U}(1)_{B-L}$ breaking \cite{Buchmuller:2012wn}. However, the core of the model --- inflation ending in a GUT-scale phase transition in combination with leptogenesis and dark matter in a supersymmetric extension of the SM --- is intrinsically tied to the GW signals discussed here.


\newpage
\begin{acknowledgments}
\textit{Acknowledgements.}
We thank Daniel Figueroa for helpful discussions on the modeling of cosmic strings
and Adeela Afzal for comments on Fig.~\ref{fig:summary}.
This project has received funding from the European's Union Horizon 2020 Research and Innovation Programme under grant agreement number 796961, 'AxiBAU' (K.\,S.).
\end{acknowledgments}


\bibliographystyle{JHEP}
\bibliography{draft}

\providecommand{\href}[2]{#2}\begingroup\raggedright\begin{thebibliography}{10}

\bibitem{Abbott:2016blz}
{\bf LIGO Scientific, Virgo} Collaboration, B.~Abbott et~al., {\it {Observation
  of Gravitational Waves from a Binary Black Hole Merger}},  {\em Phys. Rev.
  Lett.} {\bf 116} (2016), no.~6 061102,
  [\href{http://arxiv.org/abs/1602.03837}{{\tt arXiv:1602.03837}}].

\bibitem{Abbott:2016nmj}
{\bf LIGO Scientific, Virgo} Collaboration, B.~P. Abbott et~al., {\it
  {GW151226: Observation of Gravitational Waves from a 22-Solar-Mass Binary
  Black Hole Coalescence}},  {\em Phys. Rev. Lett.} {\bf 116} (2016), no.~24
  241103, [\href{http://arxiv.org/abs/1606.04855}{{\tt arXiv:1606.04855}}].

\bibitem{Abbott:2017vtc}
{\bf LIGO Scientific, VIRGO} Collaboration, B.~P. Abbott et~al., {\it
  {GW170104: Observation of a 50-Solar-Mass Binary Black Hole Coalescence at
  Redshift 0.2}},  {\em Phys. Rev. Lett.} {\bf 118} (2017), no.~22 221101,
  [\href{http://arxiv.org/abs/1706.01812}{{\tt arXiv:1706.01812}}]. [Erratum:
  Phys. Rev. Lett.121,no.12,129901(2018)].

\bibitem{Shannon:2015ect}
R.~M. Shannon et~al., {\it {Gravitational waves from binary supermassive black
  holes missing in pulsar observations}},  {\em Science} {\bf 349} (2015),
  no.~6255 1522--1525, [\href{http://arxiv.org/abs/1509.07320}{{\tt
  arXiv:1509.07320}}].

\bibitem{Kerr:2020qdo}
M.~Kerr et~al., {\it {The Parkes Pulsar Timing Array Project: Second data
  release}},  {\em Publ. Astron. Soc. Austral.} {\bf 37} (2020) e020,
  [\href{http://arxiv.org/abs/2003.09780}{{\tt arXiv:2003.09780}}].

\bibitem{Arzoumanian:2018saf}
{\bf NANOGRAV} Collaboration, Z.~Arzoumanian et~al., {\it {The NANOGrav 11-year
  Data Set: Pulsar-timing Constraints On The Stochastic Gravitational-wave
  Background}},  {\em Astrophys. J.} {\bf 859} (2018), no.~1 47,
  [\href{http://arxiv.org/abs/1801.02617}{{\tt arXiv:1801.02617}}].

\bibitem{Arzoumanian:2020vkk}
{\bf NANOGrav} Collaboration, Z.~Arzoumanian et~al., {\it {The NANOGrav
  12.5-year Data Set: Search For An Isotropic Stochastic Gravitational-Wave
  Background}},  \href{http://arxiv.org/abs/2009.04496}{{\tt
  arXiv:2009.04496}}.

\bibitem{Rajagopal:1994zj}
M.~Rajagopal and R.~W. Romani, {\it {Ultralow frequency gravitational radiation
  from massive black hole binaries}},  {\em Astrophys. J.} {\bf 446} (1995)
  543--549, [\href{http://arxiv.org/abs/astro-ph/9412038}{{\tt
  astro-ph/9412038}}].

\bibitem{Phinney:2001di}
E.~Phinney, {\it {A Practical theorem on gravitational wave backgrounds}},
  \href{http://arxiv.org/abs/astro-ph/0108028}{{\tt astro-ph/0108028}}.

\bibitem{Vaskonen:2020lbd}
V.~Vaskonen and H.~Veermäe, {\it {Did NANOGrav see a signal from primordial
  black hole formation?}},  \href{http://arxiv.org/abs/2009.07832}{{\tt
  arXiv:2009.07832}}.

\bibitem{DeLuca:2020agl}
V.~De~Luca, G.~Franciolini, and A.~Riotto, {\it {NANOGrav Hints to Primordial
  Black Holes as Dark Matter}},  \href{http://arxiv.org/abs/2009.08268}{{\tt
  arXiv:2009.08268}}.

\bibitem{Nakai:2020oit}
Y.~Nakai, M.~Suzuki, F.~Takahashi, and M.~Yamada, {\it {Gravitational Waves and
  Dark Radiation from Dark Phase Transition: Connecting NANOGrav Pulsar Timing
  Data and Hubble Tension}},  \href{http://arxiv.org/abs/2009.09754}{{\tt
  arXiv:2009.09754}}.

\bibitem{Kibble:1976sj}
T.~Kibble, {\it {Topology of Cosmic Domains and Strings}},  {\em J. Phys. A}
  {\bf 9} (1976) 1387--1398.

\bibitem{Jeannerot:2003qv}
R.~Jeannerot, J.~Rocher, and M.~Sakellariadou, {\it {How generic is cosmic
  string formation in SUSY GUTs}},  {\em Phys. Rev. D} {\bf 68} (2003) 103514,
  [\href{http://arxiv.org/abs/hep-ph/0308134}{{\tt hep-ph/0308134}}].

\bibitem{Ellis:2020ena}
J.~Ellis and M.~Lewicki, {\it {Cosmic String Interpretation of NANOGrav Pulsar
  Timing Data}},  \href{http://arxiv.org/abs/2009.06555}{{\tt
  arXiv:2009.06555}}.

\bibitem{Blasi:2020mfx}
S.~Blasi, V.~Brdar, and K.~Schmitz, {\it {Has NANOGrav found first evidence for
  cosmic strings?}},  \href{http://arxiv.org/abs/2009.06607}{{\tt
  arXiv:2009.06607}}.

\bibitem{AdvVirgo}
F.~Acernese et~al., {\it Advanced virgo: a second-generation interferometric
  gravitational wave detector},  {\em Classical and Quantum Gravity} {\bf 32}
  (2015), no.~2 024001.

\bibitem{aLIGO_era_first}
{\bf LIGO Scientific Collaboration and Virgo Collaboration} Collaboration,
  B.~P. Abbott, R.~Abbott, T.~D. Abbott, M.~R. Abernathy, F.~Acernese,
  K.~Ackley, C.~Adams, T.~Adams, P.~Addesso, R.~X. Adhikari, et~al., {\it
  Gw150914: The advanced {LIGO} detectors in the era of first discoveries},
  {\em Phys. Rev. Lett.} {\bf 116} (Mar, 2016) 131103.

\bibitem{Akutsu:2018axf}
{\bf KAGRA} Collaboration, T.~Akutsu et~al., {\it {KAGRA: 2.5 Generation
  Interferometric Gravitational Wave Detector}},  {\em Nature Astron.} {\bf 3}
  (2019), no.~1 35--40, [\href{http://arxiv.org/abs/1811.08079}{{\tt
  arXiv:1811.08079}}].

\bibitem{Auclair:2019wcv}
P.~Auclair et~al., {\it {Probing the gravitational wave background from cosmic
  strings with LISA}},  {\em JCAP} {\bf 04} (2020) 034,
  [\href{http://arxiv.org/abs/1909.00819}{{\tt arXiv:1909.00819}}].

\bibitem{Dror:2019syi}
J.~A. Dror, T.~Hiramatsu, K.~Kohri, H.~Murayama, and G.~White, {\it {Testing
  Seesaw and Leptogenesis with Gravitational Waves}},
  \href{http://arxiv.org/abs/1908.03227}{{\tt arXiv:1908.03227}}.

\bibitem{Buchmuller:2019gfy}
W.~Buchmuller, V.~Domcke, H.~Murayama, and K.~Schmitz, {\it {Probing the scale
  of grand unification with gravitational waves}},  {\em Phys. Lett.} {\bf B}
  (2020) 135764, [\href{http://arxiv.org/abs/1912.03695}{{\tt
  arXiv:1912.03695}}].

\bibitem{King:2020hyd}
S.~F. King, S.~Pascoli, J.~Turner, and Y.-L. Zhou, {\it {Gravitational waves
  and proton decay: complementary windows into GUTs}},
  \href{http://arxiv.org/abs/2005.13549}{{\tt arXiv:2005.13549}}.

\bibitem{Buchmuller:2012wn}
W.~Buchmuller, V.~Domcke, and K.~Schmitz, {\it {Spontaneous $B\!-\!L$ Breaking
  as the Origin of the Hot Early Universe}},  {\em Nucl. Phys.} {\bf B862}
  (2012) 587--632, [\href{http://arxiv.org/abs/1202.6679}{{\tt
  arXiv:1202.6679}}].

\bibitem{Buchmuller:2013lra}
W.~Buchmuller, V.~Domcke, K.~Kamada, and K.~Schmitz, {\it {The Gravitational
  Wave Spectrum from Cosmological $B-L$ Breaking}},  {\em JCAP} {\bf 1310}
  (2013) 003, [\href{http://arxiv.org/abs/1305.3392}{{\tt arXiv:1305.3392}}].

\bibitem{Leblond:2009fq}
L.~Leblond, B.~Shlaer, and X.~Siemens, {\it {Gravitational Waves from Broken
  Cosmic Strings: The Bursts and the Beads}},  {\em Phys. Rev.} {\bf D79}
  (2009) 123519, [\href{http://arxiv.org/abs/0903.4686}{{\tt
  arXiv:0903.4686}}].

\bibitem{Monin:2008mp}
A.~Monin and M.~B. Voloshin, {\it {The Spontaneous breaking of a metastable
  string}},  {\em Phys. Rev.} {\bf D78} (2008) 065048,
  [\href{http://arxiv.org/abs/0808.1693}{{\tt arXiv:0808.1693}}].

\bibitem{Monin:2009ch}
A.~Monin and M.~B. Voloshin, {\it {Destruction of a metastable string by
  particle collisions}},  {\em Phys. Atom. Nucl.} {\bf 73} (2010) 703--710,
  [\href{http://arxiv.org/abs/0902.0407}{{\tt arXiv:0902.0407}}].

\bibitem{Blanco-Pillado:2013qja}
J.~J. Blanco-Pillado, K.~D. Olum, and B.~Shlaer, {\it {The number of cosmic
  string loops}},  {\em Phys. Rev.} {\bf D89} (2014), no.~2 023512,
  [\href{http://arxiv.org/abs/1309.6637}{{\tt arXiv:1309.6637}}].

\bibitem{Martins:1995tg}
C.~Martins and E.~Shellard, {\it {String evolution with friction}},  {\em Phys.
  Rev. D} {\bf 53} (1996) 575--579,
  [\href{http://arxiv.org/abs/hep-ph/9507335}{{\tt hep-ph/9507335}}].

\bibitem{Martins:1996jp}
C.~Martins and E.~Shellard, {\it {Quantitative string evolution}},  {\em Phys.
  Rev. D} {\bf 54} (1996) 2535--2556,
  [\href{http://arxiv.org/abs/hep-ph/9602271}{{\tt hep-ph/9602271}}].

\bibitem{Martins:2000cs}
C.~Martins and E.~Shellard, {\it {Extending the velocity dependent one scale
  string evolution model}},  {\em Phys. Rev. D} {\bf 65} (2002) 043514,
  [\href{http://arxiv.org/abs/hep-ph/0003298}{{\tt hep-ph/0003298}}].

\bibitem{Gouttenoire:2019kij}
Y.~Gouttenoire, G.~Servant, and P.~Simakachorn, {\it {Beyond the Standard
  Models with Cosmic Strings}},  {\em JCAP} {\bf 07} (2020) 032,
  [\href{http://arxiv.org/abs/1912.02569}{{\tt arXiv:1912.02569}}].

\bibitem{Saikawa:2020swg}
K.~Saikawa and S.~Shirai, {\it {Precise WIMP Dark Matter Abundance and Standard
  Model Thermodynamics}},  {\em JCAP} {\bf 08} (2020) 011,
  [\href{http://arxiv.org/abs/2005.03544}{{\tt arXiv:2005.03544}}].

\bibitem{Smits:2008cf}
R.~Smits, M.~Kramer, B.~Stappers, D.~R. Lorimer, J.~Cordes, and A.~Faulkner,
  {\it {Pulsar searches and timing with the square kilometre array}},  {\em
  Astron. Astrophys.} {\bf 493} (2009) 1161--1170,
  [\href{http://arxiv.org/abs/0811.0211}{{\tt arXiv:0811.0211}}].

\bibitem{Audley:2017drz}
{\bf LISA} Collaboration, P.~Amaro-Seoane et~al., {\it {Laser Interferometer
  Space Antenna}},  \href{http://arxiv.org/abs/1702.00786}{{\tt
  arXiv:1702.00786}}.

\bibitem{LIGOScientific:2019vic}
{\bf LIGO Scientific, Virgo} Collaboration, B.~P. Abbott et~al., {\it {Search
  for the isotropic stochastic background using data from Advanced LIGO’s
  second observing run}},  {\em Phys. Rev.} {\bf D100} (2019), no.~6 061101,
  [\href{http://arxiv.org/abs/1903.02886}{{\tt arXiv:1903.02886}}].

\bibitem{ET}
{\bf Einstein Telescope} Collaboration {\em
  http://www.et-gw.eu/index.php/etsensitivities}.

\bibitem{Ade:2015xua}
{\bf Planck} Collaboration, P.~A.~R. Ade et~al., {\it {Planck 2015 results.
  XIII. Cosmological parameters}},  {\em Astron. Astrophys.} {\bf 594} (2016)
  A13, [\href{http://arxiv.org/abs/1502.01589}{{\tt arXiv:1502.01589}}].

\bibitem{Vilenkin:1982hm}
A.~Vilenkin, {\it {Cosmological evolution of monopoles connected by strings}},
  {\em Nucl. Phys.} {\bf B196} (1982) 240--258.

\end{thebibliography}\endgroup


\end{document}